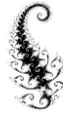

# #neverforget – Photobiomodulation Against Alzheimer's Disease: A Systematic Review

Joachim Enengl[1], Peter Dungel[*,2]

Vienna, June 6, 2019

[1]UAS Technikum Wien, Hoechstaedtplatz 6, 1200 Vienna, Austria

[*]Correspondence: peter.dungel@trauma.lbg.ac.at

[2]Ludwig Boltzmann Institute for Experimental and Clinical Traumatology, Vienna, Austria

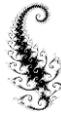

# Abstract

Alzheimer's disease affects an ever-increasing number of people in the aging population. Current treatment options are limited to a narrow time frame at the mild to moderate stage of dementia, and patients are confronted with the inevitable progression of their disease. Most investigational drugs fail to prove their efficacy in clinical trials, and there are but a few preventative measures that one can take. A novel treatment approach, using photobiomodulation to increase the brain's mitochondrial function and prevent neuronal apoptosis, has shown promising results in in vitro and in vivo experiments. This systematic review aims at providing a comprehensive summary on the available research on photobiomodulation against Alzheimer's disease to support the translation of this modality from bench to bedside. It shows that the mechanistic action has been largely understood on a cellular level, safe and effective doses have been found in animal models, and first human case studies provide reason to enter large scale clinical trials. A brief outlook on study design concludes this review and provides a basis for further work on the topic.

# Keywords

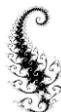

# 1 Introduction

According to the American Alzheimer's Association, 11% of older Americans (age 65 and older) suffer from Alzheimer's disease (AD) [1]. There is a similar situation in Austria, where about 100 000 people (6.4 % of Austrians aged 65 and older) suffer from AD with a prospected growth to 230 000 by the year 2050 [2]. This rise in numbers is mainly caused by the aging population, with a prolonged life expectancy and the generation of "Baby-Boomers" (people born in the 1950s and 1960s) as driving factors [3]. Affecting such a large number of individuals, Alzheimer's disease is a challenge on the health care system. About one billion Euro per year are spent on AD-care in Austria alone [2,4], three quarters of which are non-medical expenses related to personal assistance in activities of daily life. Aside from being costly, AD also claims its toll on the quality of life of caregivers. Compared to people who aren't affected by AD directly or indirectly, those who care for Alzheimer's patients have shown signs of a weaker immune system, more pro-inflammatory proteins, as well as significantly shorter telomeres [5]. As of yet there is no medical treatment that can either cure, prevent, or completely stop the progression of AD. The lack of treatment options is aggravated by a lack of reliable diagnostic tools that could detect the disease before the onset of symptoms. The pre-symptomatic period of Alzheimer's disease may last for decades, but once symptoms start to show, there is only a small therapeutic window before neurodegeneration renders the affected person completely dependent on external care.

The key to a cure of the disease may still lay in the dark, but there is an increasing amount of research being done that suggests that literal enlightenment might pave the way against AD. Starting off by simply searching for "low level light therapy Alzheimer's disease" at the PubMed database of the NCBI and using the references within relevant publications, it quickly becomes obvious that there are many different names for treatments utilizing light in the UV-VIS-NIR range. While Kendric C. Smith suggested us to use the "simple and correct term, phototherapy" [6], the scientific consensus recently introduced the term photobiomodulation (PBM) for what used to be called low level laser/light therapy (LLLT) [7,8]. As the literature also quickly shows, it seems possible to apply photobiomodulation to the brain transcranially, for which even more different terms and abbreviations exist. Figure 1 provides an overview to the multiple terms describing the same technique.

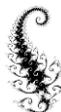

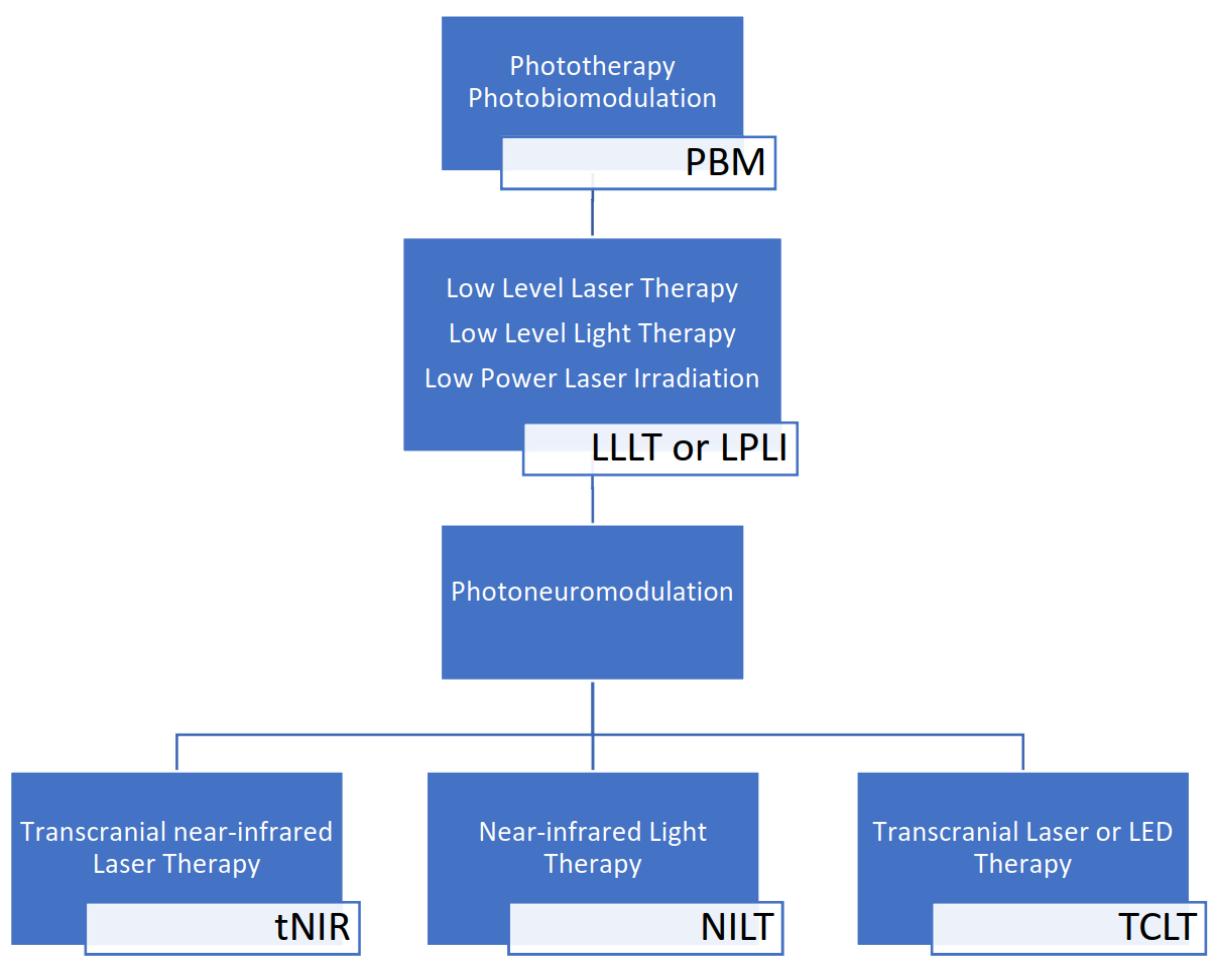

Figure 1: There are many names used for the use of light in medical applications, which can all be grouped under the umbrella term of photobiomodulation. Used against Alzheimer's disease the method utilizes comparatively low power, or low level, near-infrared laser or LED light which can be applied in a transcranial manner for photo-neuro-modulation [9].

As Table 1 shows, reducing the literature search to the MeSH terms "Photobiomodulation" and "Alzheimer's disease" falls a long way short of what can be found when performing a query containing a combination of all search terms from Figure 1 and identifiers for Alzheimer's disease and dementia. There was just one obvious mismatch when using the broader search term and it was easy to identify when "NILT" was found to be used as an abbreviation for "Northern Ireland Life and Times".

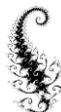

Table 1: Comparison of search results in the NCBI PubMed database from a free search containing multiple and historic names for photobiomodulation versus MeSH terms only.

|  | Combination of terms from Figure 1 and an identifier for Alzheimer's disease | MeSH Terms for AD and PBM |
|---|---|---|
| Total number of results | 36 | 8 |
| Species: Humans | 16 | 4 |
| Type: Review article | 5 | 0 |
| Obvious mismatch | 1 | 0 |

The 5 Review articles [10–14] come to the conclusion, that photobiomodulation with low level red to near infrared light offers a compelling method against neurodegeneration.

# 2 Literature Summary

Photobiomodulation (PBM) is defined as the use of monochromatic or quasi-monochromatic light from a low power laser or LED source to modify or modulate biological functions. Such a modulation is based on the presence of chromophores in cells and tissues. These chromophores are molecules capable of absorbing light, whose excitation can influence further molecules and biochemical pathways with the potential for a therapeutic effect [15].

The target molecule of PBM, mainly discussed in literature, lies within Complex IV (cytochrome c oxidase) of the mitochondrial respiration system. Photons that interfere with the cytochrome oxidase enzyme in the NIR frequency range are absorbed, essentially acting as an exogenous supply of highly energized electrons to the mitochondrial respiratory chain. Following this immediate, primary effect, the increase in cytochrome oxidase activity triggers secondary effects by activating enzymatic pathways which are coupled to the mitochondria's role in energy metabolism, cell homeostasis, and cell survival signaling. The secondary effects of photobiomodulation amplify the primary effects by upregulating the levels of cytochrome oxidase and thus creating more of the target molecule for PBM [13,16].

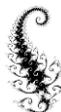

The energy density delivered by PBM is generally too low to cause concerns about heating and tissue destruction [17], and it appears to be safe even when applied long term [18]. As a treatment option for head and neck pain, for treatments of arthritis, and for carpal tunnel syndrome, PBM was shown to have an efficacy beyond the placebo effect, [19], which lead to the first FDA-approved low-level light therapy devices for pain relief [20].

Interest in the use of photobiomodulation for neuro-rehabilitation is growing and it has already shown potential in treating traumatic brain injury, stroke, psychiatric disorders, as well as neurodegenerative disease in general [21–23].

## 1.1 Transcranial Photobiomodulation with Red to Near Infra-Red Light

To make PBM work on the brain, one can benefit from the optical tissue window which allows wavelengths of light between approximately 650 nm and 1200 nm to travel through skin and skull, i.e. transcranially. The boundaries of the optical tissue window are defined by the strong absorption of hemoglobin and water. The penetration depth of wavelengths within those boundaries was measured for example by Tedford et al, who achieved a maximal penetration depth at a wavelength of 808 nm [24,25]. Wang and Li evaluated this further and confirmed 810 nm as well as 660 nm to be the best suitable wavelengths for transcranial photobiomodulation [26].

Considering an illumination of the brain at even deeper depths without having to open the skull, Bungart et al suggest the use of nanoparticles as an alternative light source for PBM [27]. These nanoparticles, termed 'Bioluminescence Resonance Energy Transfer to Quantum Dots' (BRET-Qdots), emit light in the NIR wavelength range when their luciferase enzyme is activated with coelenterazine-h substrate. The reported downside to a

Enengl, Dungel #neverforget – PBM vs AD: A Systematic Review Page **6** of **25**

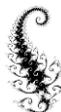

nanoparticle treatment are concerns about the toxicity of the heavy metal content of the BRET-Qdots, which are yet to be fully alleviated.

# 1 Molecular Mechanisms of Photobiomodulation and their Implications for Alzheimer's Disease

The impact of photobiomodulation on cell culture and animal models of Alzheimer's disease has been extensively studied. A summary of the progress on each stage of development from basic research to preclinical and clinical Phase I and II trials is presented to form the base for a treatment plan.

### 2..1.1 PBM and AD in vitro

Sommer et al report that photobiomodulation reduces amyloid beta (A$\beta_{25-35}$) aggregates in human neuroblastoma cells [28] and Yang et al observed a reduction in amyloid beta induced oxidative and inflammatory responses in rat primary cortical astrocytes [29]. Looking further at the implications of PBM on neuro-inflammatory pathways, Song et al observed the effects of He-Ne laser light (632.8 nm) at doses from 3 J/cm² to 50 J/cm² on central nervous system resident macrophages (Microglia) in human SH-SY5Y neuronal cells and conclude that 20 J/cm² at intervals of 24 h was the optimal dose to attenuate cell death by reducing microglia-mediated neurotoxicity via the Src/Syk signaling pathways [30].

Liang et al and Zhang et al examined the pathways leading to cell apoptosis following A$\beta_{25-35}$ aggregation on neuronally differentiated PC12 cells and observed that a treatment with low power light produced with a He-Ne Laser at 623.8 nm demonstrated a positive effect of PBM [31–33]. Zhang et al specifically report that PBM activates protein kinase B (PKB / Akt) at a dose of 2 J/cm², which in turn promotes a series of anti-apoptotic effects such as the inhibition of translocation of the pro-apoptotic Yes-associated protein (YAP) from the cytoplasm to the nucleus [33]. PBM with a He-Ne Laser also appears to activate protein kinase C (PKC), thereby affecting downstream apoptotic proteins in a dose dependent manner [32]. Irradiation for 5 to 20 min with an energy density of 0.52 mW/cm², corresponding to a delivered dose from 0.156 J/cm² to 0.624 J/cm², decreases levels of the pro-apoptotic Bax and either increases or does not affect levels of the anti-apoptotic Bcl-$X_L$ as compared to control cells with Aβ$_{25-35}$ aggregation. This effect was reversed when an irradiation time of 40 minutes with 0.52 mW/cm² was reached, corresponding to a delivered dose of 1.248 J/cm², which illustrates the dose dependence of PBM when used to counteract Aβ-induced cell apoptosis. The reports from Liang et al from their use of a 632.8 nm He-Ne Laser at a dose of 2 J/cm² on neuronal PC12 cells with Aβ$_{25-35}$ aggregation state that PBM has a pro-survival effect by acting on the Akt/GSK3b/b-catenin pathway [31]. Building on to the hypothesis that amyloid-beta induced neurotoxicity and dendrite atrophy may be a consequence of brain-derived neurotropic factor (BDNF) deficiency, Meng et al report that ERK can be photo-activated with a 632.8 nm He-Ne laser, delivering doses from 0.5 J/cm² to 4 J/cm² applied to Aβ$_{25-35}$-treated human SH-SY5Y cells, which upregulates BDNF in a CREB-dependent manner [34]. A summary of the cell signaling effects of photobiomodulation described to this point is shown in Figure 2.

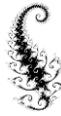

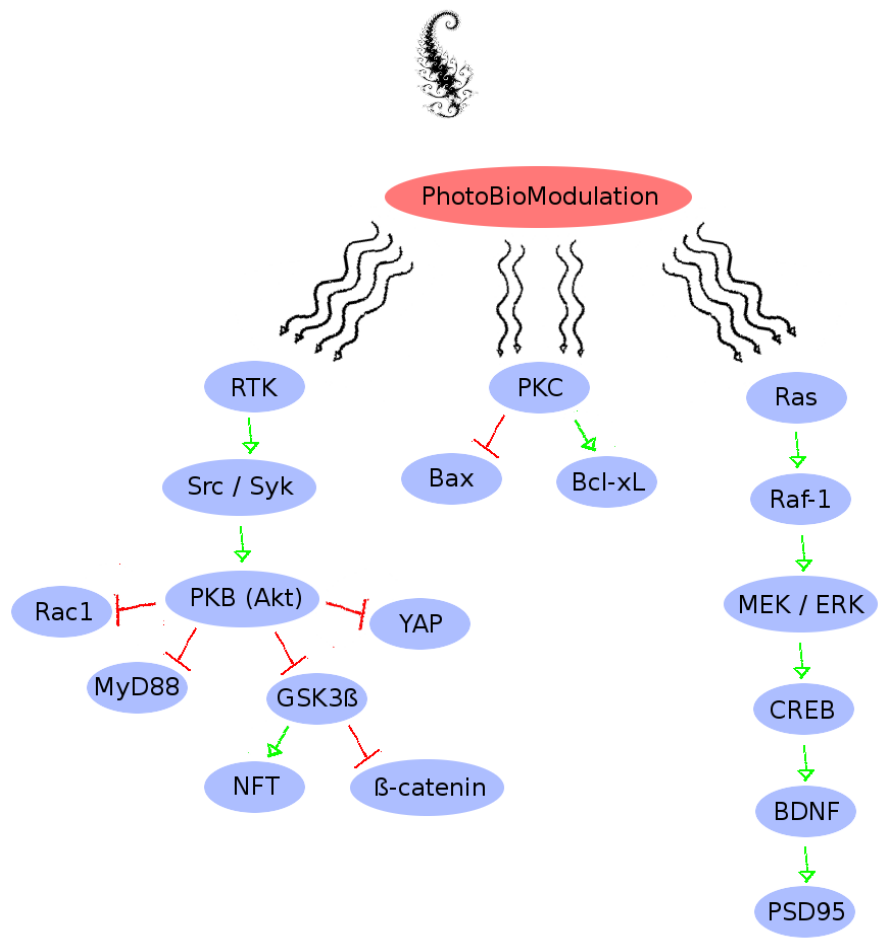

*Figure 2: Summary of the signaling pathways influenced by photobiomodulation, which appear to inhibit Aβ-induced nerve cell apoptosis while simultaneously promoting nerve cell survival.*

### 2..1.2 PBM and AD in vivo

Using a mouse model of Alzheimer's disease, Farfara et all and Oron et al have shown to ameliorate disease progression by stimulating the proliferation of mesenchymal stem cells (MSCs) with photobiomodulation [35,36]. Weekly treatments of PBM with a dose of 1 J/cm² applied to the bone marrow of AD-mice iappears to increase the MSCs ability to phagocytose Aβ-proteins in the brain, which leads to improved cognitive capacity and spatial learning after a total treatment duration of 2 months, compared to a sham-treated control group. This is an interesting observation, albeit outside the scope of this review, since it builds not on the effect of PBM on braincells directly, but rather secondary effects through stimulation of MSCs.

Using two transgenic mouse models for AD, engineered to either develop neurofibrillary tangles (NFT) or Aβ-plaques, and treating them for a total of 20 times over the course of four weeks with LED light at 670 nm, delivering a dose of 4 J/cm² per treatment, Purushothuman et al demonstrated histochemical evidence of a beneficial effect [37]. NFTs, hyperphosphorylated tau-protein, and oxidative stress markers were reduced to near wildtype levels and PBM treatment also reduced Aβ-plaques in both number and in size.

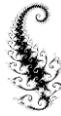

De Taboada et al used a GaAIAs diode laser with a wavelength of 808 nm ± 10 nm to deliver increasing doses, from 4.8 J/cm² (1.2 J/cm² at cortical surface) to 48 J/cm² (12 J/cm² at cortical surface), three times per week for a total duration of six months on transgenic mice engineered to develop Aβ-plaques [38]. Their results show an attenuation of amyloid development, leading to the conclusion that an early and regular administration of PBM has the potential to halt progression from MCI to AD.

## 2 Applications of Transcranial Photobiomodulation on Human Subjects Supporting its Therapeutic Value for Neurological Use

In an opinion article by Gonzalez-Lima and Barret [17], the authors summarize how the development of photobiomodulation (PBM) has evolved over the last 40 years to a point where it is beginning to be used for cognitive-enhancing applications.

They built a placebo controlled study focusing on the beneficial cognitive and emotional effects in humans upon this opinion, and already two weeks after a single therapy session a significant improvement was observed [39]. These beneficial effects are reflected in improvements in reaction time in a sustained-attention psychomotor vigilance task (PVT), in a delayed match-to-sample memory task (DMS), and a self-reported Positive and Negative Affect Schedule (PANAS-X). Blanco et al then used the Wisconsin Card Sorting Task (WCST) directly after receiving the PBM treatment to assess the effects of transcranial infrared laser stimulation on executive function in a placebo controlled study on healthy human participants [40]. Here, low level laser light at a wavelength of 1064 nm and power density of 250 mW/cm² was applied to two locations on the right portion of the forehead for a total of four minutes per site. The total supplied dose of this study again corresponds to the ones stated above, being 60 J/cm² per site, and the treatment group performed significantly better than the control group during a WCST performed.

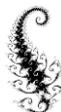

Schiffer et al report beneficial effects of transcranial PBM on anxiety and depression after a study in which PBM with an LED array with a peak wavelength at 810 nm and an energy density of also 250 mW/cm² was used [41]. They calculated that from the supplied dose of 60 J/cm² at the scalp, an effective 2.1 J/cm² (3.7 %) reached the dura of the brain. Positive effects of a single treatment session compared to a placebo control group were observed two weeks after PBM, which waned off by the time of a further assessment after four weeks.

As stated by Hashimi et al in their article about the role of PBM in neurorehabilitation [21], transcranial photobiomodulation with low level red to near infrared light has already been used on human subjects with moderate Alzheimer's disease, although no peer reviewed publications exist as of yet. This is slowly changing, with new devices being developed and more individuals being treated [42,43].

## 1.2 Minimum Effective and Maximum Tolerated Dose

When red-to-near-infrared monochromatic or quasi-monochromatic light is supplied to the cytochrome oxidase in neuronal mitochondria, the photons from this light influence the proton gradient of the mitochondrial membrane and thus modulate ATP production. The effective dose of PBM appears to follow the principle of hormesis, meaning that low doses are stimulatory while higher doses are less effective or even counterproductive. The minimum effective and maximum tolerated dose has been experimentally evaluated in vitro and in vivo, the translation of these results to transcranial application on humans is supported by results from studies on transmission factors and penetration depth of red and NIR light in intermediate tissue.

## 3 Hormesis

The hermetic effect of PBM was shown by Gonzalez-Lima and Barrett in a study on rats, where 660 nm LED arrays with a power density of 9 mW/cm² showed the highest increase in cytochrome oxidase activity at a dose of 10.9 J/cm² (13.6 % increase), slightly less at the higher dose of 21.6 J/cm² (10.3 % increased activity) and finally a point of only non-significant activity increase (3 %) at the highest dose of 32.9 J/cm² [17]. Using fluorescence-quenching to measure neuronal oxygen consumption by looking at oxygen concentration in the cortex, Rojas et al report a 5 % increase in oxygen consumption at 1 J/cm² and a 16 % increase at a dose of 5 J/cm² when using PBM with 660 nm and an energy density of 9 mW/cm² on rats [44].

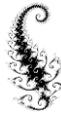

## 4    Transmission

Tedford et al quantitatively analyzed the amount of light that reaches the brain by using human cadaver heads, which they sectioned in such way that all of the intermittent tissues were kept intact and the optical energy was measured "on site" within the brain [24]. Results show an exponential increase in fluence rates with a linear decrease in distance between measurement probe and light source. An absorption coefficient for each of the measured wavelengths is also provided which can then be used to calculate the dose supplied to any point inside the brain using equation (1):

$$F(x)\left[\frac{mW}{cm^2}\right] = F_0\left[\frac{mW}{cm^2}\right] \cdot e^{-\mu[cm^{-1}] \cdot x[cm]}$$

With F(x) being the radiant intensity delivered to the point inside the brain in mW/cm² per 1 mW/cm² delivered to the surface of the scalp ($F_0$), and x being the distance between the surface of the scalp and the point of interest inside the brain. The value for µ is the absorption coefficient of the tissue at each supplied wavelength; at 660 nm it is 3.3504 cm$^{-1}$, at 808 nm it is 2.5541 cm$^{-1}$, and at 940 nm it is 3.3922 cm$^{-1}$.

Lychagov et al studied the NIR transmittance of the human skull and demonstrated that 0.5 % to 5 % of the emitted light from a 1 W laser with a wavelength of 810 nm is transmitted in transcranial application, proving the exponential relationship between the thickness of skin and bone and light transmission in the NIR-range [45].

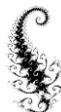

Comparing the transmission of red light at 633 nm to that of near infrared light at 830 nm in a human cadaveric model, Jagdeo et all observed that 830 nm has a superior transmittance to 633 nm and an occipital application allows the highest amount of energy to enter the brain, as Table 2 shows [46].

*Table 2: Transmittance measurements of 830 nm and 633 nm light through human cadaver heads at multiple positions. The values stated in this table are an excerpt from the results published in [46].*

|  | Control (10 mm of air) | Behind skin and skull | | |
|---|---|---|---|---|
|  |  | temporal | frontal | occipital |
| 830 nm | 33.3 mW/cm² | 0.3 mW/cm² | 0.71 mW/cm² | 3.9 mW/cm² |
|  | 100% | 0.9% | 2.1% | 11.7% |
| 633 nm | 67.5 mW/cm² | <0.001 mW/cm² | 0.37 mW/cm² | 0.44 mW/cm² |
|  | 100% | 0.0% | 0.5% | 0.7% |

Bungart et al investigated different wavelengths for transcranial PBM and came to the conclusion that a combination of 660 nm and 810 nm provides superior penetration compared to a combination of 980 nm and 1064 nm [26].

Once the light reaches the inside of the skull, the depth of penetration is limited by the optical properties of the brain tissue. Table 3 shows an excerpt of the results from Stolik et al's ex vivo, post mortem measurements of the penetration depths of red and near infrared light in brain tissue, which suggest that light with longer wavelength will penetrate deeper into the tissue [47].

*Table 3: Optical penetration depth of human brain tissue for different wavelengths in the red to near infrared spectrum of light. The values stated in this table are an excerpt from the results published in [47]*

| Wavelength | 632.8 nm | 675 nm | 780 nm | 835 nm |
|---|---|---|---|---|

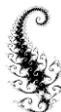

| Penetration Depth [mm ± standard error] n = 10 | 0.92 ± 0.08 | 1.38 ± 0.13 | 2.17 ± 0.16 | 2.52 ± 0.19 |
|---|---|---|---|---|

To investigate which wavelength within the optical tissue window will penetrate the deepest and deliver the highest dose, the values for the tissue absorption coefficient as well as the penetration depth were extracted from [24] and [47] and extrapolated by use of polynomial trend lines as shown in Figure 3.

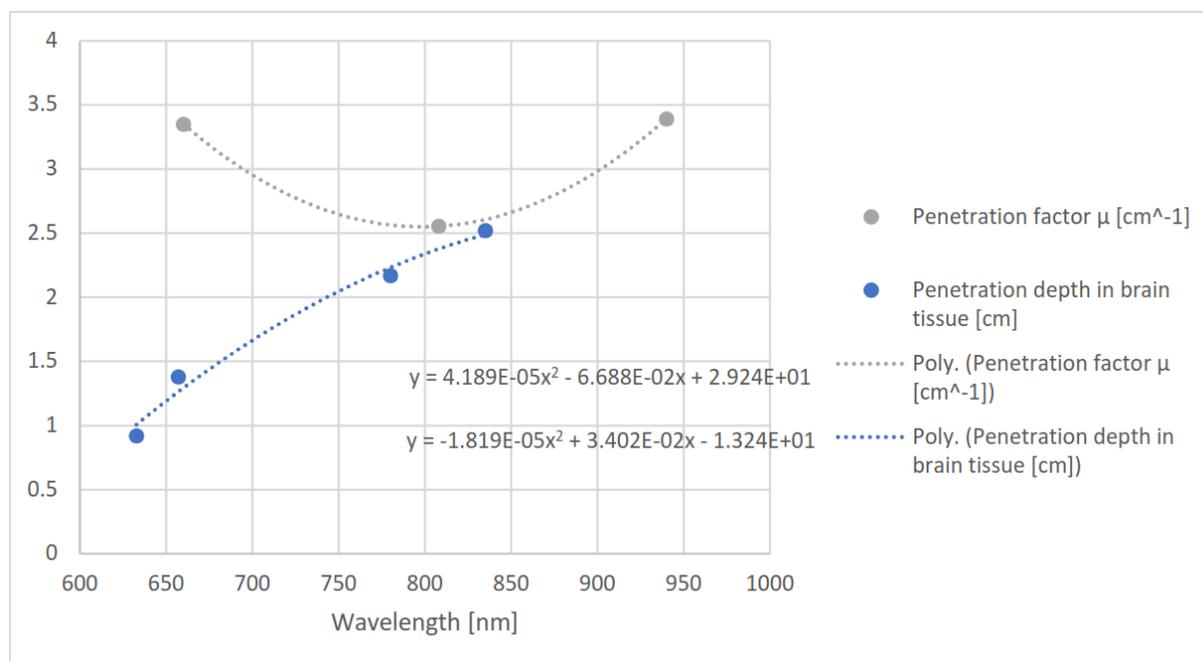

*Figure 3: Using the literature reports on the transmittance of near infrared light through the scalp and skull, as well as the penetration depth inside the brain, mathematical trends are established in order to use meaningful values when calculating the supplied dose of wavelengths that are not covered by literature [9].*

While Figure 3 suggest that a wavelength between 700 nm and 830 nm is likely to have the most effective and also the deepest delivery of light energy to areas inside the brain, an article by Wu et al claims that illumination with 730 nm and 980 nm appears to be ineffective when used as a modality for traumatic brain injury in mice and they suggest to use a mix of 665 nm and 810 nm instead [48].

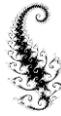

# 5  Energy Density and Dosage

Summing up the available body of research on PBM to the brain, an application time long enough to achieve a dose in the minimum effective range between 5 – 10 J/cm² is needed on a neuronal level, so, considering the exponential energy decrease during transcranial application, energy densities of 60 J/cm² applied to the human scalp are suggested. With an energy density of 250 mW/cm² this dose can be delivered within 4 minutes while PBM with a device supplying 100 mW/cm² would be able to deliver the required dose in 10 minutes.

## 1.3  Outlook on a Study Design

As a randomized, controlled, clinical intervention study, the focus needs to be on individuals who are at particular, specific risk to develop Alzheimer's disease (AD), or at an early stage with evidence of AD neuropathology [49]. A stagnation of the MMSE-score of the treatment group can be chosen as the most easily detectable primary endpoint of the study. It is expected that the MMSE-score of the control group falls by 2 – 5 points per year [50].

To reach sufficient power and significance of the study, at least 15 participants per group will be required when assuming that the MMSE-scores of treatment– and control group differ by 4 points at the end of a year-long trial. Even with a relatively large standard deviation of 5 points per group this can result in a relatively large effect size of 0.8 with a level of significance of 0.1 and power of 0.8. Since Jacova et al have shown that in longitudinal studies on dementia a dropout rate of up to 55 % can be expected after 2 years [51], this should be taken into account by recruiting twice as many participants than were calculated to be sufficient, resulting in a minimum n = 30 per group.

To achieve a higher sensitivity for any changes during the treatment protocol it is suggested to supplement the standard MMSE by a more sophisticated neuropsychological test battery. One such option could be the CERAD-NP-Test Battery, which includes the Mini-Mental State Examination (MMSE), word fluency testing, the Modified Boston Naming Test (MBNT), constructive practical testing, as well as learning, repeating, and recognizing of a word list

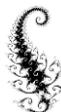

[52–61]. The pathophysiological assessment with MRI at the start of the study is useful to support the diagnosis of early AD. A second MRI at the end of the treatment period might provide more useful in combination with a sensitive algorithm that can reliably detect a change in brain atrophy, for example the automated MRI assessment by Desikan et al which has already demonstrated its usefulness as a diagnostic marker for mild cognitive impairment (MCI) and AD [62].

# 3 Conclusion

When comparing the specifications of PBM devices used in the available literature, especially wavelength and intensity, a very inhomogeneous picture is created. One thing that is clear, however, is that the optical window for deep tissue penetration suggests best results at wavelengths in the far red and near infrared range, e.g. at 810 nm. The irradiation intensity of such a device needs to allow delivery of a dose of up to 60 J/cm² to the scalp within a reasonable period of time while providing an energy density below the risk of thermal damage. The amalgamation of the available literature as presented in this systematic review allows the assumption that photobiomodulation with the right settings can be a useful tool in the fight against Alzheimer's disease. Study protocols for large placebo-controlled studies are readily available and we should be able to get significant results with representative statistical power within a year. The time to act is now – lest we forget.

# 4 Acknowledgements

This study was supported by FFG Basisprogramm grant #853128.

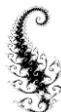

# 5 References

1. Alzheimer's Association. 2014 Alzheimer's Disease Facts and Figures [Internet]. 225 N. Michigan Ave., Fl.17, Chicago, IL 60601-7633: Alzheimer's Association; 2014. Available from: http://www.alz.org/downloads/Facts_Figures_2014.pdf

2. Österreichische Alzheimergesellschaft. Alzheimer: Zahlen & Statistik [Internet]. 2015 [cited 2015 Nov 10]. Available from: http://www.alzheimer-gesellschaft.at/index.php?id=46

3. Statistik Austria. Österreich. Zahlen. Daten. Fakten [Internet]. Statistik Austria; 2015 May. Available from: http://www.statistik.at/wcm/idc/idcplg?IdcService=GET_NATIVE_FILE&dDocName=029266

4. Dal-Bianco P. DFP Literaturstudium: M. Alzheimer - State of the Art. ÖÄZ. Österreichische Gesellschaft für Neurologie; 2010. p. 40–53.

5. Turkington C, Mitchell DR. The encyclopedia of Alzheimer's disease. 2nd ed. New York: Facts On File; 2010.

6. Smith KC. Laser (and LED) therapy is phototherapy. Photomed Laser Ther [Internet]. 2005 [cited 2015 Dec 7];23:78–80. Available from: http://online.liebertpub.com/doi/pdf/10.1089/pho.2005.23.78

7. Arany PR. Photobiomodulation: Poised from the Fringes. Photomed Laser Surg [Internet]. 2012 [cited 2018 Oct 5];30:507–9. Available from: https://www.liebertpub.com/doi/10.1089/pho.2012.9884

8. Anders JJ, Lanzafame RJ, Arany PR. Low-Level Light/Laser Therapy Versus Photobiomodulation Therapy. Photomed Laser Surg [Internet]. 2015 [cited 2018 Oct 5];33:183–4. Available from: https://www.liebertpub.com/doi/10.1089/pho.2015.9848

9. Enengl J. Low-Level Laser/Light Therapy in a Multidisciplinary Approach on Alzheimer Rehabilitation [Internet] [Master Thesis | Biomedical Engineering Sciences]. [Vienna, Austria]: UAS Technikum Wien; 2016. Available from: https://permalink.obvsg.at/AC13410415

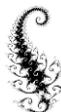

10. de la Torre JC. Treating cognitive impairment with transcranial low level laser therapy. J Photochem Photobiol B [Internet]. 2017;168:149–55. Available from: http://www.sciencedirect.com/science/article/pii/S1011134416306765

11. de la Torre JC. Cerebral perfusion enhancing interventions: A new strategy for the prevention of Alzheimer dementia. Brain Pathol [Internet]. 2016 [cited 2016 Jul 4]; Available from: http://doi.wiley.com/10.1111/bpa.12405

12. Santana-Blank L, Rodríguez-Santana E, Santana-Rodríguez KE, Reyes H. "Quantum Leap" in Photobiomodulation Therapy Ushers in a New Generation of Light-Based Treatments for Cancer and Other Complex Diseases: Perspective and Mini-Review. Photomed Laser Surg [Internet]. 2016 [cited 2018 Oct 13];34:93–101. Available from: https://www.liebertpub.com/doi/10.1089/pho.2015.4015

13. Gonzalez-Lima F, Barksdale BR, Rojas JC. Mitochondrial respiration as a target for neuroprotection and cognitive enhancement. Biochem Pharmacol [Internet]. 2014 [cited 2015 Nov 25];88:584–93. Available from: http://linkinghub.elsevier.com/retrieve/pii/S0006295213007417

14. Lapchak PA. Transcranial near-infrared laser therapy applied to promote clinical recovery in acute and chronic neurodegenerative diseases. Expert Rev Med Devices [Internet]. 2012 [cited 2015 Dec 9];9:71–83. Available from: http://informahealthcare.com/doi/abs/10.1586/erd.11.64

15. Rojas JC, Gonzalez-Lima F. Neurological and psychological applications of transcranial lasers and LEDs. Biochem Pharmacol [Internet]. 2013 [cited 2015 Nov 26];86:447–57. Available from: http://linkinghub.elsevier.com/retrieve/pii/S0006295213003833

16. Hayworth CR, Rojas JC, Padilla E, Holmes GM, Sheridan EC, Gonzalez-Lima F. In Vivo Low-level Light Therapy Increases Cytochrome Oxidase in Skeletal Muscle. Photochem Photobiol [Internet]. 2010 [cited 2015 Nov 26];86:673–680. Available from: http://onlinelibrary.wiley.com/doi/10.1111/j.1751-1097.2010.00732.x/full

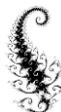

17. Gonzalez-Lima F, Barrett DW. Augmentation of cognitive brain functions with transcranial lasers. Front Syst Neurosci [Internet]. 2014 [cited 2015 Nov 26];8. Available from: http://journal.frontiersin.org/article/10.3389/fnsys.2014.00036/abstract

18. McCarthy TJ, De Taboada L, Hildebrandt PK, Ziemer EL, Richieri SP, Streeter J. Long-Term Safety of Single and Multiple Infrared Transcranial Laser Treatments in Sprague–Dawley Rats. Photomed Laser Surg [Internet]. 2010 [cited 2015 Nov 25];28:663–7. Available from: http://www.liebertonline.com/doi/abs/10.1089/pho.2009.2581

19. Naeser MA. Photobiomodulation of pain in carpal tunnel syndrome: review of seven laser therapy studies. Photomed Laser Ther [Internet]. 2006 [cited 2015 Nov 25];24:101–110. Available from: http://online.liebertpub.com/doi/abs/10.1089/pho.2006.24.101

20. Fulop AM, Dhimmer S, Deluca JR, Johanson DD, Lenz RV, Patel KB, et al. A meta-analysis of the efficacy of laser phototherapy on pain relief. Clin J Pain [Internet]. 2010 [cited 2015 Dec 3];26:729–736. Available from: http://journals.lww.com/clinicalpain/Abstract/2010/10000/A_Meta_analysis_of_the_Efficacy_of_Laser.12.aspx

21. Hashmi JT, Huang Y-Y, Osmani BZ, Sharma SK, Naeser MA, Hamblin MR. Role of Low-Level Laser Therapy in Neurorehabilitation. PM&R [Internet]. 2010 [cited 2015 Nov 25];2:S292–305. Available from: http://linkinghub.elsevier.com/retrieve/pii/S1934148210012530

22. Naeser MA, Hamblin MR. Potential for Transcranial Laser or LED Therapy to Treat Stroke, Traumatic Brain Injury, and Neurodegenerative Disease. Photomed Laser Surg [Internet]. 2011 [cited 2015 Nov 25];29:443–6. Available from: http://www.liebertonline.com/doi/abs/10.1089/pho.2011.9908

23. Chang J, Ren Y, Wang R, Li C, Wang Y, Ping Chu X. Transcranial Low-Level Laser Therapy for Depression and Alzheimer's Disease. Neuropsychiatry [Internet]. 2018 [cited 2018 Oct 5];08. Available from: http://www.jneuropsychiatry.org/peer-review/transcranial-lowlevel-laser-therapy-for-depression-and-alzheimers-disease-12428.html

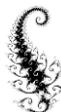

24. Tedford CE, DeLapp S, Jacques S, Anders J. Quantitative analysis of transcranial and intraparenchymal light penetration in human cadaver brain tissue. Lasers Surg Med [Internet]. 2015 [cited 2015 Nov 26];47:312–22. Available from: http://doi.wiley.com/10.1002/lsm.22343

25. Tedford CE, DeLapp S, Jacques S, Anders J. Re: "Quantitative analysis of transcranial and intraparenchymal light penetration in human cadaver brain tissue" Lasers in Surgery and Medicine, 2015;47(4):312-322. Lasers Surg Med [Internet]. 2015 [cited 2018 Oct 5];47:466–466. Available from: http://doi.wiley.com/10.1002/lsm.22377

26. Wang P, Li T. Which wavelength is optimal for transcranial low-level laser stimulation? J Biophotonics [Internet]. 2018 [cited 2018 Oct 6];0:e201800173. Available from: https://doi.org/10.1002/jbio.201800173

27. Bungart BL, Dong L, Sobek D, Sun GY, Yao G, Lee JC-M. Nanoparticle-emitted light attenuates amyloid-β-induced superoxide and inflammation in astrocytes. Nanomedicine Nanotechnol Biol Med [Internet]. 2014 [cited 2015 Nov 25];10:15–7. Available from: http://linkinghub.elsevier.com/retrieve/pii/S1549963413005832

28. Sommer AP, Bieschke J, Friedrich RP, Zhu D, Wanker EE, Fecht HJ, et al. 670 nm Laser Light and EGCG Complementarily Reduce Amyloid-β Aggregates in Human Neuroblastoma Cells: Basis for Treatment of Alzheimer's Disease? Photomed Laser Surg [Internet]. 2012 [cited 2015 Nov 25];30:54–60. Available from: http://online.liebertpub.com/doi/abs/10.1089/pho.2011.3073

29. Yang X, Askarova S, Sheng W, Chen JK, Sun AY, Sun GY, et al. Low energy laser light (632.8 nm) suppresses amyloid-β peptide-induced oxidative and inflammatory responses in astrocytes. Neuroscience [Internet]. 2010 [cited 2015 Nov 25];171:859–68. Available from: http://linkinghub.elsevier.com/retrieve/pii/S030645221001273X

30. Song S, Zhou F, Chen WR. Low-level laser therapy regulates microglial function through Src-mediated signaling pathways: implications for neurodegenerative diseases. J Neuroinflammation [Internet]. 2012 [cited 2015 Dec 9];9:1–17. Available from: http://www.biomedcentral.com/content/pdf/1742-2094-9-219.pdf

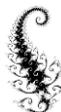

31. Liang J, Liu L, Xing D. Photobiomodulation by low-power laser irradiation attenuates Aβ-induced cell apoptosis through the Akt/GSK3β/β-catenin pathway. Free Radic Biol Med [Internet]. 2012 [cited 2015 Nov 25];53:1459–67. Available from: http://linkinghub.elsevier.com/retrieve/pii/S0891584912004923

32. Zhang L, Xing D, Zhu D, Chen Q. Low-Power Laser Irradiation Inhibiting Aβ25-35-induced PC12 Cell Apoptosis via PKC Activation. Cell Physiol Biochem [Internet]. 2008;22:215–22. Available from: http://www.karger.com/DOI/10.1159/000149799

33. Zhang H, Wu S, Xing D. Inhibition of Aβ25–35-induced cell apoptosis by Low-power-laser-irradiation (LPLI) through promoting Akt-dependent YAP cytoplasmic translocation. Cell Signal [Internet]. 2012 [cited 2015 Nov 25];24:224–32. Available from: http://linkinghub.elsevier.com/retrieve/pii/S089865681100283X

34. Meng C, He Z, Xing D. Low-Level Laser Therapy Rescues Dendrite Atrophy via Upregulating BDNF Expression: Implications for Alzheimer's Disease. J Neurosci [Internet]. 2013 [cited 2015 Oct 10];33:13505–17. Available from: http://www.jneurosci.org/cgi/doi/10.1523/JNEUROSCI.0918-13.2013

35. Farfara D, Tuby H, Trudler D, Doron-Mandel E, Maltz L, Vassar RJ, et al. Low-Level Laser Therapy Ameliorates Disease Progression in a Mouse Model of Alzheimer's Disease. J Mol Neurosci [Internet]. 2015 [cited 2015 Dec 20];55:430–6. Available from: http://link.springer.com/10.1007/s12031-014-0354-z

36. Oron A, Oron U. Low-Level Laser Therapy to the Bone Marrow Ameliorates Neurodegenerative Disease Progression in a Mouse Model of Alzheimer's Disease: A Minireview. Photomed Laser Surg [Internet]. 2016 [cited 2016 Jul 5]; Available from: http://dx.doi.org/10.1089/pho.2015.4072

37. Purushothuman S, Johnstone DM, Nandasena C, Mitrofanis J, Stone J. Photobiomodulation with near infrared light mitigates Alzheimer's disease-related pathology in cerebral cortex–evidence from two transgenic mouse models. Alzheimer's Res Ther [Internet]. 2014 [cited 2015 Dec 9];6. Available from: http://www.biomedcentral.com/content/pdf/alzrt232.pdf

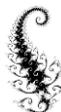

38. De Taboada L, Yu J, El-Amouri S, Gattoni-Celli S, Richieri S, McCarthy T, et al. Transcranial Laser Therapy Attenuates Amyloid-β Peptide Neuropathology in Amyloid-β Protein Precursor Transgenic Mice. J Alzheimer's Dis [Internet]. 2011 [cited 2015 Dec 9];23:521–535. Available from: http://www.researchgate.net/profile/Luis_De_Taboada/publication/253420093_Transcranial_laser_therapy_alters_amyloid_precursor_protein_processing_and_improves_mitochondrial_function_in_a_mouse_model_of_Alzheimer's_Disease/links/54b7e54f0cf2c27adc476a6e.pdf

39. Barrett DW, Gonzalez-Lima F. Transcranial infrared laser stimulation produces beneficial cognitive and emotional effects in humans. Neuroscience [Internet]. 2013 [cited 2015 Nov 26];230:13–23. Available from: http://linkinghub.elsevier.com/retrieve/pii/S0306452212011268

40. Blanco NJ, Maddox WT, Gonzalez-Lima F. Improving executive function using transcranial infrared laser stimulation. J Neuropsychol [Internet]. 2015; Available from: http://dx.doi.org/10.1111/jnp.12074

41. Schiffer F, Johnston AL, Ravichandran C, Polcari A, Teicher MH, Webb RH, et al. Psychological benefits 2 and 4 weeks after a single treatment with near infrared light to the forehead: a pilot study of 10 patients with major depression and anxiety. Behav Brain Funct [Internet]. 2009 [cited 2015 Nov 26];5:46. Available from: http://behavioralandbrainfunctions.biomedcentral.com/articles/10.1186/1744-9081-5-46

42. Saltmarche AE, Naeser MA, Ho KF, Hamblin MR, Lim L. Significant Improvement in Cognition in Mild to Moderately Severe Dementia Cases Treated with Transcranial Plus Intranasal Photobiomodulation: Case Series Report. Photomed Laser Surg [Internet]. 2017 [cited 2018 Oct 13];35:432–41. Available from: https://www.liebertpub.com/doi/10.1089/pho.2016.4227

43. Berman MH, Halper JP, Nichols TW, H J, Lundy A, Huang JH. Photobiomodulation with Near Infrared Light Helmet in a Pilot, Placebo Controlled Clinical Trial in Dementia Patients Testing Memory and Cognition. J Neurol Neurosci [Internet]. 2017 [cited 2018 Oct 13];08.

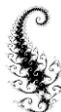

Available from: http://www.jneuro.com/neurology-neuroscience/photobiomodulation-with-near-infrared-light-helmet-in-a-pilot-placebo-controlled-clinical-trial-in-dementia-patients-testing-memor.php?aid=18528

44. Rojas JC, Bruchey AK, Gonzalez-Lima F. Low-level light therapy improves cortical metabolic capacity and memory retention. J Alzheimers Dis JAD. 2012;32:741–52.

45. Lychagov VV, Tuchin VV, Vilensky MA, Reznik BN, Ichim T, De Taboada L. Experimental study of NIR transmittance of the human skull. In: Tuchin VV, editor. Proc SPIE 6085 Complex Dyn Fluct Biomed Photonics III [Internet]. 2006 [cited 2015 Dec 9]. Available from: http://proceedings.spiedigitallibrary.org/proceeding.aspx?articleid=1273934

46. Jagdeo JR, Adams LE, Brody NI, Siegel DM. Transcranial Red and Near Infrared Light Transmission in a Cadaveric Model. Hamblin M, editor. PLoS ONE [Internet]. 2012 [cited 2015 Dec 9];7:e47460. Available from: http://dx.plos.org/10.1371/journal.pone.0047460

47. Stolik S, Delgado JA, Perez A, Anasagasti L. Measurement of the penetration depths of red and near infrared light in human "ex vivo" tissues. J Photochem Photobiol B [Internet]. 2000 [cited 2015 Dec 9];57:90–93. Available from: http://www.sciencedirect.com/science/article/pii/S1011134400000828

48. Wu Q, Xuan W, Ando T, Xu T, Huang L, Huang Y-Y, et al. Low-Level Laser Therapy for Closed-Head Traumatic Brain Injury in Mice: Effect of Different Wavelengths. Lasers Surg Med [Internet]. 2012 [cited 2018 Oct 9];44:218–26. Available from: http://doi.wiley.com/10.1002/lsm.22003

49. BOKDE A. The Road Ahead to Cure Alzheimer's Disease: Development of Biological Markers and Neuroimaging Methods for Prevention Trials Across all Stages and Target Populations. 2014 [cited 2015 Nov 25]; Available from: http://edepositireland.ie/handle/2262/74864

50. Tombaugh T. Test-retest reliable coefficients and 5-year change scores for the MMSE and 3MS. Arch Clin Neuropsychol [Internet]. 2005 [cited 2016 Jan 19];20:485–503. Available from: http://acn.oxfordjournals.org/cgi/doi/10.1016/j.acn.2004.11.004

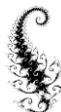

51. Jacova C, Hsiung G-YR, Feldman HH. Dropouts and refusals in observational studies: Lessons for prevention trials. Neurology [Internet]. 2006;67:S17–20. Available from: http://www.neurology.org/content/67/9_suppl_3/S17.abstract

52. Mirra SS, Heyman A, McKeel D, Sumi SM, Crain BJ, Brownlee LM, et al. The Consortium to Establish a Registry for Alzheimer's Disease (CERAD): Part II. Standardization of the neuropathologic assessment of Alzheimer's disease. Neurology [Internet]. 1991;41:479–479. Available from: http://www.neurology.org/content/41/4/479.abstract

53. MIRRA SS, GEARING M, MCKEEL DWJ, CRAIN BJ, HUGHES JP, BELLE GV, et al. Interlaboratory Comparison of Neuropathology Assessments in Alzheimer's Disease: A Study of the Consortium to Establish a Registry for Alzheimer's Disease (CERAD). J Neuropathol Exp Neurol [Internet]. 1994;53:303–15. Available from: http://journals.lww.com/jneuropath/Fulltext/1994/05000/Interlaboratory_Comparison_of_Neuropathology.12.aspx

54. Patterson MB, Mack JL, Mackell JA, Thomas R, Tariot P, Weiner M, et al. A Longitudinal Study of Behavioral Pathology Across Five Levels of Dementia Severity in Alzheimer's Disease: The CERAD Behavior Rating Scale for Dementia. Alzheimer Dis Assoc Disord [Internet]. 1997;11:40–4. Available from: http://journals.lww.com/alzheimerjournal/Fulltext/1997/00112/A_Longitudinal_Study_of_Behavioral_Pathology.6.aspx

55. Tariot PN. CERAD behavior rating scale for dementia. Int Psychogeriatr [Internet]. 1997 [cited 2016 Jan 19];8:317–320. Available from: http://journals.cambridge.org/abstract_S1041610297003542

56. Satzger W, Hampel H, Padberg F, Bürger K, Nolde T, Ingrassia G, et al. Zur praktischen Anwendung der CERAD-Testbatterie als neuropsychologisches Demenzscreening. Nervenarzt [Internet]. 2001 [cited 2016 Jan 19];72:196–203. Available from: http://link.springer.com/article/10.1007/s001150050739

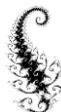

57. Chandler MJ, Lacritz LH, Hynan LS, Barnard HD, Allen G, Deschner M, et al. A total score for the CERAD neuropsychological battery. Neurology [Internet]. 2005 [cited 2016 Jan 19];65:102–106. Available from: http://www.neurology.org/content/65/1/102.short

58. Karrasch M, Sinerva E, Gronholm P, Rinne J, Laine M. CERAD test performances in amnestic mild cognitive impairment and Alzheimer's disease. Acta Neurol Scand [Internet]. 2005 [cited 2016 Jan 19];111:172–9. Available from: http://doi.wiley.com/10.1111/j.1600-0404.2005.00380.x

59. für die AgeCoDe Study Group, Luck T, Riedel-Heller SG, Wiese B, Stein J, Weyerer S, et al. CERAD-NP-Testbatterie: Alters-, geschlechts- und bildungsspezifische Normen ausgewählter Subtests: Ergebnisse der German Study on Ageing, Cognition and Dementia in Primary Care Patients (AgeCoDe). Z Für Gerontol Geriatr [Internet]. 2009 [cited 2016 Jan 19];42:372–84. Available from: http://link.springer.com/10.1007/s00391-009-0031-y

60. Rossetti HC, Munro Cullum C, Hynan LS, Lacritz LH. The CERAD Neuropsychologic Battery Total Score and the Progression of Alzheimer Disease: Alzheimer Dis Assoc Disord [Internet]. 2010 [cited 2016 Jan 19];24:138–42. Available from: http://content.wkhealth.com/linkback/openurl?sid=WKPTLP:landingpage&an=00002093-201004000-00005

61. Schönknecht ODP, Hunt A, Toro P, Guenther T, Henze M, Haberkorn U, et al. Bihemispheric cerebral FDG PET correlates of cognitive dysfunction as assessed by the CERAD in Alzheimer's disease. Clin EEG Neurosci [Internet]. 2011 [cited 2016 Jan 19];42:71–76. Available from: http://eeg.sagepub.com/content/42/2/71.short

62. Desikan RS, Cabral HJ, Hess CP, Dillon WP, Glastonbury CM, Weiner MW, et al. Automated MRI measures identify individuals with mild cognitive impairment and Alzheimer's disease. Brain [Internet]. 2009 [cited 2015 Nov 25];132:2048–57. Available from: http://www.brain.oxfordjournals.org/cgi/doi/10.1093/brain/awp123